\documentclass[sigconf]{acmart}

\usepackage{booktabs} 

\setcopyright{none}

\begin{document}
\title[Genre classification by automatic identification of repeating patterns]{On large-scale genre classification in symbolically encoded music by automatic identification of repeating patterns}

\author{Andres Ferraro}
\orcid{}
\affiliation{%
  \institution{Music Technology Group - Universitat Pompeu Fabra}
  \streetaddress{Roc Boronat, 138}
  \city{Barcelona}
  \state{Spain}
  \postcode{08018}
}
\affiliation{%
  \institution{Universidad de la Rep\'ublica, Montevideo, Uruguay}
  \streetaddress{}
  \city{}
  \state{}
  \postcode{}
}
\email{andres.ferraro@upf.edu}

\author{Kjell Lemstr\"om}
\affiliation{%
  \institution{University of Helsinki, Department of Computer Science}
  \streetaddress{Ernst Lindelöfin katu 1}
  \city{Helsinki}
  \state{Finland}
  \postcode{00560}}
\email{kjell.lemstrom@helsinki.fi}

\begin{abstract}
The importance of repetitions in music is well-known. In this paper, we study music repetitions in the context of effective and efficient automatic genre classification in large-scale music-databases. We aim at enhancing the access and organization of pieces of music in Digital Libraries by allowing automatic categorization of entire collections by considering only their musical content. We handover to the public a set of genre-specific patterns to support research in musicology. The patterns can be used, for instance, to explore and analyze the relations between musical genres.

There are many existing algorithms that could be used to identify and extract repeating patterns in symbolically encoded music. In our case, the extracted patterns are used as representations of the pieces of music on the underlying corpus and, consecutively, to train and evaluate a classifier to automatically identify genres. In this paper, we apply two very fast algorithms enabling us to experiment on large and diverse corpora. Thus, we are able to find patterns with strong discrimination power that can be used in various applications. We carried out experiments on a corpus containing over 40,000 MIDI files annotated with at least one genre. The experiments suggest that our approach is scalable and capable of dealing with real-world-size music collections.
\end{abstract}

%
%



\maketitle

\section{Introduction}

Since repetitions play such an important role in most of the musical genres \cite{margulis2014repeat}, in this paper we aim at automatically identify repetitions that are intrinsic for a genre. We call a (melodic) sequence of notes that repeats in a piece of music a {\bf pattern}. A repetition of a pattern may be transposed by a constant pitch-shift, i.e., transposition invariance applies.

The detection of repeated patterns in music collections has been studied in several areas of Music Information Retrieval (MIR). One of the first MIR problems was to retrieve a piece of a music in a music database based on a given note-sequence query-pattern \cite{Kosugi:2000:PQS:354384.354520,lem,2657,orio2006music,1285862,raffel2015large,ROCAMORA2014272} as summarized in \cite{INR-042}.

Another conventional MIR task is the problem of automatic genre classification \cite{Sturm2014}. Accurate genre identification would enable both effective navigation in music collections and plausible music recommendations. In literature, one can find genre classification algorithms both to audio and symbolic music data. In this paper, we focus on symbolic music because of the straightforward processing and robust, efficient and effective pattern detection algorithms.

Applied techniques for symbolic-music genre-classification \\(SMGC) have been manifold. The classification has been based on meta-level features, such as note duration and musical key distributions, and on identifying repeated note patterns using different features and representations; see e.g. \cite{CORREA2016190}. To our knowledge, SMGC experiments with the largest database to date was made by Li et al. \cite{li2006factored}. Their database contained  14,000 folk songs, but they did not yield adequate performance results, such as the area under curve (AUC) for the receiver operating characteristic (ROC)\cite{6984840}. Furthermore, to our knowledge there has been no study considering multiple labels on a corpus. In this paper, we use a very large corpus that includes a high number of genres. For the experiments, we form the database of those MIDI files that are annotated in the corpus with at least one genre. The resulting set of patterns extracted by our algorithms is not restricted to any specific genre or tradition.

The existence of specific repeating patterns intrinsic to a genre is well-known. For instance, \cite{odekerken} studied the relation between Jazz and Ragtime, evolving to a later work to identify repeating patterns in a Ragtime corpus \cite{koops}. Evaluations of several algorithms identifying patterns in a Folk music was given in \cite{boot}, where music was classified based on the compression given by the extracted patterns. 

In this paper, we apply two algorithms, SIA \cite{doi:10.1076/jnmr.31.4.321.14162} and P2 \cite{Ukkonen:2}, originally developed to solve slightly different problems. They both work with polyphonic music represented geometrically as points in a Euclidean space. SIA was originally developed to discover recurring patterns, P2 to efficiently find occurrences of a query pattern in a corpus. SIA and P2 run in $O(n^2\log n)$ and $O(mn\log m)$ time, respectively, where $n$ and $m$ represent the number of notes in the corpus and the number of notes in the query pattern. 

Janssen et al. \cite{Janssen2014} give a review on pattern discovery in music and discuss the challenges of the task. According to them, one of the main problems is that the experiments in the literature are carried out with different datasets, which makes it practically impossible to compare the results. In this paper, we mitigate this problem by using a public dataset that comprehends multiple genres and styles. 

We publish for the community all the extracted patterns in association with pieces of music they refer to, to contribute and facilitate further studies in musicology. This dataset allows performing studies like Monson's \cite{monson1999riffs}, but in a manner that scales up.


\section{Data}

\begin{table}
 \begin{center}
 \begin{tabular}{lcc}
  \toprule
  \# dataset & files with annotation & total annotations \\
  \midrule
  MASD & 17785 & 24623 \\
  MAGD & 23496 & 37237 \\
  top-MAGD & 22535 & 34867 \\
  \bottomrule
 \end{tabular}
\end{center}
 \caption{Number of files with genre annotations in Lakh.}
 \label{annotations}
\end{table}

In this work, we use the {\it Lakh dataset} \cite{raffel} of MIDI files that is mapped to the Million Song Dataset (MSD) \cite{Bertin-Mahieux2011}. There are multiple datasets with annotations of genres for the MSD. We chose 3 subsets of annotations presented by Schindler et al \cite{Schindler:3}. For each subset in the Lakh dataset, Table~\ref{annotations} gives the number of files that contain at least one annotation of genre (having mapped them to MSD) and the total number of annotations. 

We use the same names from Schidler et al. work for the subsets of annotations: The annotations were originally extracted from the All Music Guide\footnote{\url{http://allmusic.com}}, hence we call the subset containing genre annotations \textbf{MSD Allmusic Genre Dataset}, or {\bf MAGD} for short, and the subset containing style annotations {\bf MASD}. The styles contained in MASD are given in Table~\ref{masd}. The third subset, called {\bf top-MAGD}, is the subset of MAGD that includes only the top 13 genres shown in Table~\ref{top-magd}.  

The difference between the datasets, as stated by the authors in \cite{Schindler:3}, is that MASD attempts to distinguish the songs into different sub-genres.


\begin{table}
 \begin{center}
 \begin{tabular}{lc}
  \toprule
  Genre & Number of songs \\
  \midrule
Pop/Rock & 21024\\
Electronic & 3460\\
Country & 2410\\
R\&B & 2040\\
Jazz & 1179\\
Latin & 1410\\
International & 1008\\
Rap & 701\\
Vocal & 698\\
New Age & 496\\
Folk & 200\\
Reggae & 141\\
Blues & 100\\
  \hline
  Total & 34867\\
  \bottomrule
 \end{tabular}
\end{center}
 \caption{Songs for a genre annotated in top-MAGD.}
 \label{top-magd}
\end{table}

\begin{table}
 \begin{center}
 \begin{tabular}{lc}
  \toprule
  Style & Number of songs \\
  \midrule
Big Band & 362 \\
Blues Contemporary & 114\\
Country Traditional & 2065\\
Dance & 2017\\
Electronica & 605\\
Experimental & 733\\
Folk International & 707\\
Gospel & 405\\
Grunge Emo & 302\\
Hip Hop Rap & 801\\
Jazz Classic & 496\\
Metal Alternative & 978\\
Metal Death & 214\\
Metal Heavy & 282\\
Pop Contemporary & 4291\\
Pop Indie & 1147\\
Pop Latin & 838\\
Punk & 113\\
Reggae & 127\\
RnB Soul & 544\\
Rock Alternative & 700\\
Rock College & 977\\
Rock Contemporary & 2890\\
Rock Hard & 2096\\
Rock Neo Psychedelia & 519\\
\hline
Total & 24623\\
\bottomrule
 \end{tabular}
\end{center}
 \caption{Songs for a style annotated in MASD.}
 \label{masd}
\end{table}

\section{Patterns extracted from MIDI files}

We processed all the tracks of the MIDI files with SIA and P2 algorithms, extracting patterns containing at least 3 notes (shorter were considered musically meaningless). Moreover, we applied some further filtering on the found patterns. For the SIA algorithm we used the following filtering thresholds: 
\begin{itemize}
\item Length: meaningful patterns must have at least 3 notes
\item Compactness: the relative length of the pattern with respect to the length of the whole piece of music
\item Temporal density: the more notes in a given time frame the higher the temporal density
\end{itemize}

The patterns passing the first threshold (length) were considered for the second filtering round where a combination of the two remaining thresholds was used (compactness and temporal density). We conducted experiments on 4 combinations, as shown in Table~\ref{siatonic}

As algorithm P2 searches for occurrences of a given query sequence within a piece of music, we first segmented each piece in the database in subsections with different lengths and overlaps and, then, used these subsections as query sequences for P2. For each occurrence found, P2 returns a similarity value (between 0 and 1). In Table~\ref{p2tonic}, we show the thresholds used for the following 3 parameters:
\begin{itemize}
\item Length: number note events
\item Offset: the number of intervening elements allowed
\item Similarity: given by P2
\end{itemize}

For both algorithms the thresholds were selected based on preliminary tests on a smaller music corpus. To decide which thresholds combinations were to be used, we looked at the number of returned patterns: too restrictive values does not return patterns and too permissive thresholds return too many patterns. 

\begin{table}
 \begin{center}
 \begin{tabular}{lcc}
  \toprule
  Name & Compactness & Temporal Density \\
  \midrule
  $Sia-1$ & 0.7 & 0.05 \\
  $Sia-2$ & 0.4 & 0.05 \\
  $Sia-3$ & 0.4 & 0.25 \\
  $Sia-4$ & 0.7 & 0.25 \\
  \bottomrule
 \end{tabular}
\end{center}
 \caption{Threshold combinations used for the SIA algorithm to filter the patterns.}
 \label{siatonic}
\end{table}

\begin{table}
 \begin{center}
 \begin{tabular}{lccc}
  \toprule
  Name & Length (notes) & Offset (notes) & Similarity  \\
  \midrule
  $P2-3$ & 3 & 2 & 0.9 \\
  $P2-4$ & 4 & 2 & 0.9 \\
  $P2-5$ & 5 & 3 & 0.5 \\
  $P2-8$ & 8 & 3 & 0.5 \\
  $P2-10$ & 10 & 3 & 0.5 \\
  $P2-15$ & 15 & 3 & 0.5 \\
  \bottomrule
 \end{tabular}
\end{center}
 \caption{Considered threshold combinations for the P2 algorithm to filter the patterns.}
 \label{p2tonic}
\end{table}

To our knowledge, there is no large-scale public-dataset of patterns for multiple genres. Therefore, we evaluate the goodness of the extracted patterns indirectly by measuring the capability of the algorithms in performing genre or style classification for the pieces of music. Should the used configuration give a good classification, the extracted patterns capture important information on the genre/style of the considered piece of music.  

\begin{table*}
 \begin{center}
 \begin{tabular}{lccccccc}
 \toprule
  & \multicolumn{3}{c}{\textbf{topMAGD}} & \multicolumn{3}{c}{\textbf{MASD}} & \\
  
  \textbf{Name} & \textbf{AUC ROC} & \textbf{F1 measure}& \textbf{Accuracy}&  \textbf{AUC ROC} & \textbf{F1 measure}& \textbf{Accuracy} & \textbf{\#patterns}\\
  
  \midrule
  \textbf{$Sia-1$ }& 0.749 (0.014) & 0.628 (0.003) & 0.484 (0.005) & 0.761 (0.004)& 0.455 (0.003)& 0.342 (0.009)& 130394 \\
  \textbf{$Sia-2$} & 0.753 (0.014)& 0.637 (0.003) & 0.500 (0.006) & 0.760 (0.004)& 0.464 (0.003)& 0.358 (0.009)& 236586 \\
  \textbf{$Sia-3$} & 0.750 (0.014)& 0.624 (0.004) & 0.463 (0.006)& 0.757 (0.005)& 0.443 (0.004)& 0.321 (0.007)& 81547 \\
  \textbf{$Sia-4$} & 0.745 (0.013)& 0.618 (0.004) & 0.447 (0.009)& 0.755 (0.005)& 0.431 (0.004)& 0.298 (0.007)& 54890  \\
  \textbf{$P2-3$} & 0.740 (0.007)& 0.636 (0.005)& 0.547 (0.008)& 0.738 (0.013) & 0.451 (0.004) & 0.318 (0.007) & 1002232 \\
  \textbf{$P2-4$} & 0.769 (0.005)& 0.662 ( 0.007) & 0.620 (0.008) & 0.771 (0.009)& 0.468 (0.007)& 0.316 (0.008)& 2083114 \\
  \textbf{$P2-5$} & \textbf{0.816} (0.005)& 0.649 (0.005) & 0.641 (0.005)& \textbf{0.815} (0.009)& 0.431 (0.010)& 0.270 (0.006)& 2763773 \\
  \textbf{$P2-8$ }& 0.805 (0.003)& 0.626 (0.006)& 0.632 (0.005) & 0.809 (0.005)& 0.374 (0.009)& 0.226 (0.007)& 3560508 \\
  \textbf{$P2-10$} & 0.785 (0.002)& 0.619 (0.006)& 0.629 (0.006)& 0.793 (0.005)& 0.355 (0.009)& 0.211 (0.005)& 3693755 \\
  \textbf{$P2-15$} & 0.750 (0.003)& 0.613 (0.008)& 0.626 (0.007) & 0.757 (0.005)& 0.334 (0.008)& 0.196 (0.006)& 2671666 \\
  \bottomrule
 \end{tabular}
\end{center}
 \caption{SIA and P2 on genre classification using patterns on top-MAGD and MASD datasets. The performance measured with AUC-ROC, F1-score and Accuracy. The column \#patterns indicates the total number of distinct patterns identified.}
 \label{tonic_results}
\end{table*}

\subsection{Converting note patterns into common representation}

In MIDI files, the temporal resolution is indicated in the header. This is known as ticks per quarter note (TPQN). The  higher the TPQN, the higher the resolution.

In order to be able to compare the extracted patterns across all the MIDI files, we need to convert them to a common representation. To this end, we used the information of TPQN on each file and converted them to a common resolution. For the common TPQN, we used a low value (six) to make sure that sequences with small differences fall in the same representation.

Initially, we extracted the patterns in the MIDI files and represented them as sequences of tuples \verb=(position|tone)=, where the position is the temporal indicator of the corresponding note and tone the pitch of the note. For instance, a pattern

\begin{itemize}
\item[] \begin{verbatim}(0|0)  (545|3)  (682|10)  (818|12)\end{verbatim}
\end{itemize}
was converted to common TPQN (originally ticks per quarter note was 480):

\begin{itemize}
\item[] \begin{verbatim}(0|0)(6|3)(8|10)(10|12) \end{verbatim}
\end{itemize}

\subsection{Classification of genre with extracted patterns}

Measuring the genre / style classification performance in the case of multiple labels for a piece of music is not an easy task. To this end, we follow the approach by \cite{oramas2017multi}, who measured the area under the ROC curve (AUC ROC). We also consider some extra measures complementing AUC, that is, the F1 measure and the averaged accuracy of each class.

As shown in Tables~\ref{top-magd} and~\ref{masd}, the genres in topMAGD dataset are much more imbalanced than the styles in MASD dataset. This challenge mimics nicely the ones found in a real world applications. 

For the classification, we create a matrix where the columns correspond to the extracted patterns and the rows to the pieces of music in our database. Each cell of the matrix counts the occurrences of the respective pattern within the corresponding piece of music. The resulting matrix is subsequently given as input to the classification algorithm for which we use logistic regression and weights of the classes that are automatically balanced by the algorithm. We use the scikit-learn implementation \cite{scikit-learn} for this task. In order to avoid overfitting we use a 5-fold cross-validation.

For each possible configuration, we repeat independently the same classification process for the annotations with the topMAGD dataset and the MASD dataset. As it can be seen in Table~\ref{tonic_results}, in the AUC ROC columns of each dataset \textbf{$P2-5$} gives the best performance for the task. This setting outperforms also the accuracy reported by the authors of the dataset \cite{Schindler:3}. They used the same dataset but different features and classifiers.

When observing the two algorithms individually, there is no remarkable difference in the performance between the different configurations for the SIA algorithm, but changing the configuration of P2 would have a rather notable difference in the performance. This might suggest that the patterns extracted by different settings of SIA are more homogeneous than the ones extracted by different settings of P2. 

\section{Conclusions and future work}

We harnessed two algorithms, originally designed to different tasks, SIA and P2, for detecting patterns that allow us to automatically identify genre for a collection of MIDI songs. In our experiments, P2 gave the best patterns for the genre / style classification task.

Iterating P2 over all possible substrings of the underlying dataset generates the patterns more efficiently, giving a remarkable speedup to the process. Nevertheless, their results for the two tasks were surprisingly different in finding the set of patterns that describe well-enough the considered genres. In doing that, we noticed that P2 gives a better performance than SIA.


One could experiment also on applying other similarity algorithms, such as dynamic time warping, for the pattern detection tasks and find that a combination of the algorithms would give the best classification result. In such a case, however, it would be beneficial to apply some unsupervised learning method. Moreover, it would be interesting to apply unsupervised learning for clustering the patterns in order to analyze the relations between distinct genres or between artists.

We encourage further research on this and closely related topics by setting all our code and all the extracted patterns publicly available~\footnote{\url{https://github.com/andrebola/patterns-genres}}. 

Our future plans include building an interface to explore and query the extracted patterns and the relations that were relevant in identifying the genres. With such an interface, it would be easier for musicologists to further explore and analyze the results of this study, for instance.

\begin{acks}

The research was partly supported by the Agencia Nacional de Investigaci\'on e Innovaci\'on under the code POS\_NAC\_2016\_1\_130162 and PEDECIBA.

\end{acks}

\bibliographystyle{ACM-Reference-Format}
\bibliography{sample-bibliography}

\end{document}